\documentclass[11pt,twoside]{article}

\usepackage{asp2006}
\usepackage{epsf}
\usepackage{psfig}
\usepackage{lscape}

\begin{document}
\title{Simulating CCD images of elliptical galaxies}
\author{M. Cantiello\altaffilmark{1,2}, G. Raimondo\altaffilmark{2}, E. Brocato\altaffilmark{2}, 
J.P. Blakeslee\altaffilmark{1}, M. Capaccioli\altaffilmark{3}}   
\altaffiltext{1}{Dep. of Physics and Astronomy, Washington State University, Pullman, WA 99164}
\altaffiltext{2}{INAF-Oss. Astronomico di Teramo, Via M. Maggini, 64100, Teramo, Italy}    
\altaffiltext{3}{INAF-Oss. Astronomico di Capodimonte, Vicolo Moiariello 16, 80131, Napoli, Italy}

\begin{abstract} 
We introduce a procedure developed by the ``Teramo Stellar Populations
Tools'' group (Teramo-SPoT), specifically optimized to obtain
realistic simulations of CCD images of elliptical galaxies.

Particular attention is devoted to include the Surface Brightness
Fluctuation (SBF) signal observed in ellipticals and to simulate the
Globular Cluster (GC) system in the galaxy, and the distribution of
background galaxies present in real CCD frames. In addition to the
physical properties of the simulated objects - galaxy distance and
brightness profile, luminosity function of GC and background galaxies,
etc. -  the tool presented allows the user to set some of the main
instrumental properties - FoV, zero point magnitude, exposure
time, etc.

\end{abstract}



The light coming from distant galaxies includes a specific SBF signal,
essentially correlated with the properties of the host stellar system
(Tonry \& Schneider 1988). The existence of these luminosity
fluctuations is due to the statistical correlation between adjacent
galaxy regions (pixels).  Since its introduction, the SBF method has
been used as a reliable distance indicator for elliptical galaxies
(e.g. Tonry et al. 2001) and, more recently, as a tracer of stellar
population properties (e.g. Cantiello et al. 2003, 2005; Raimondo et
al. 2005, R05).

In order to derive SBF magnitudes from CCD images of elliptical
galaxies, very high quality CCD data are required. We have developed a
tool to simulate CCD images of elliptical galaxies including the SBF
signal and other properties of the galaxy - surface brightness
profile, distance, color profiles, contamination of background galaxies,
etc.

Due to its statistical nature, a reliable simulation of the SBF signal
needs: i) to accurately reproduce the details of the statistics
governing the stellar SBF signal, and ii) to take into account the
presence of any other source of fluctuations.  To include SBF signal
in the simulations we use the Teramo-SPoT Single-burst Stellar
Populations (SSP) models (R05, visit also the SPoT website
www.oa-teramo.inaf.it/SPoT). These models are provided by computing a
number N$_{sim}$ of independent SSP simulations for a large range of
ages and chemical compositions. The latter property of SPoT SSP models
is at the base of our simulations of realistic galaxies: we start
with a galaxy having an analytic Sersic r$^{1/n}$ profile, then, for
each pixel [i,j] at the radius $r*$ we substitute the analytic
magnitude profile $\mu_{th}(r*)_{[i,j]}$ with the integrated magnitudes
$\mu_{sim}$ as evaluated in one of the N$_{sim}$ independent SSP
simulations, having assumed $\langle \mu_{sim} \rangle
=\mu_{th}(r*)$. In this way the poissonian correlation between adjacent
pixels is introduced, preserving the galaxy brightness profile.

To be realistic, the simulation must also include the presence of GC
and background galaxies - which, in addition, can strongly affect the
fluctuations signal derived from the CCD.  These sources are indeed
included into our simulations according to their typical luminosity
functions, i.e., the total luminosity function is assumed to be the
sum of a power law for galaxies, and a gaussian distribution for the
GC component. The characteristic parameters of these functions can be
arbitrarily set by the user.  Once the luminosity functions are
randomly populated all the background galaxies are randomly
distributed in the frame, while the GC spatial distribution is
additionally convolved with an inverse power law centered on the
galaxy.


Finally, a uniform sky value is included, and the detector noise is
added according to the readout-noise and gain values of the selected
instrument.

After the galaxy profile - including SBF -, the GC system, the
background galaxies, and the detector noise properties have been
properly chosen, the simulation can be carried out. The panels of
Figure 1 show the frames associated to some of the steps described 
above, the final frame simulated, and the luminosity functions of
GC and background galaxies. 
\begin{figure}[!ht]
\plotone{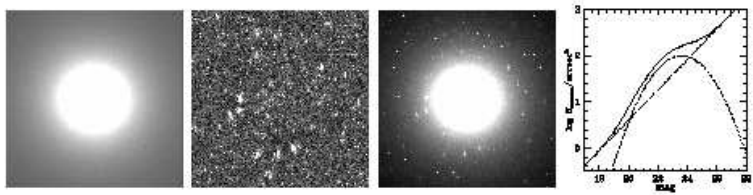}
\caption{The first three panels of the figure (left to right) show the
profile of the galaxy simulated, the frame of GC and background
galaxies, and the sum of the previous two frames plus sky and noise,
respectively. ACS camera properties are adopted for the instrumental
characteristics.  The last plot shows the luminosity functions adopted
of the GC (short dashed line) and the background galaxies (long dashed
line), and their sum (solid line).}
\end{figure}


The capabilities of the procedure here briefly described makes it
useful to simulate astronomical data for a wide range of applications.
As a specific case we mention the use of this tool to simulate
realistic runs at defined telescopes with the aim of measuring SBF.
For example, we have applied this technique to simulate ISAAC@VLT
Ks-band, and ACS@HST F814W-band (reported Figure 1) images of given
ellipticals in order to evaluate the proper exposure times required to
reach a defined S/N ratio for objects at different distances.


\end{document}